# Ferromagnetic Resonance of Co/Gd and Co/Tb Multilayers

**S. Demirtas[1], I. Harward[1], R. E. Camley[1] and Z. Celinski[1]**

*[1]Department of Physics, University of Colorado at Colorado Springs,*

*Colorado Springs, CO 80918 USA*

**M. R. Hossu[2] and A. R. Koymen[2]**

*[2]Department of Physics, University of Texas at Arlington, Arlington, TX 76019 USA*

**C. Yu[3] and M. J. Pechan[3]**

*[3]Department of Physics, Miami University, Oxford, OH 45056 USA*

***Abstract***

The in-plane dynamics of ferrimagnetic Co/Gd multilayers are investigated by means of ferromagnetic resonance, magneto-optical Kerr effect and SQUID magnetometry. The power absorbed from these multilayers is strongly temperature dependent. For example, the resonant peak for a (Co 40 Å /Gd 40 Å)$_8$ multilayer vanishes approximately 50 K below room temperature. We have further investigated Gd/Co/Gd and Tb/Co/Tb trilayers with different thicknesses of Gd (5-7 Å), Tb (1-7 Å) and Co (30-40 Å). At room temperature, these Co-based trilayers show a shift of approximately 600 Oe at 24 GHz in the uniform ferromagnetic resonance field, compared to pure Co film, indicating the exchange coupling between the Co and Gd. The shift in the field for the resonance increases as the temperature is decreased. Furthermore the resonance linewidth increases as the temperature is decreased. The experimental results are in good agreement with our theoretical calculations.

## Introduction

Magnetic multilayers have been an area of active research for a considerable period, partly because their properties, such as giant magnetoresistance, have lead to applications in the recording industry. In addition to the common multilayers involving transition metal ferromagnets with nonmagnetic spacers, there has also been a substantial study of transition metal / rare earth multilayers where both materials are ferromagnetic. These systems exhibit multiple magnetic configurations and phase transitions between these states at different external magnetic fields and temperatures.

The Co/Gd and Fe/Gd multilayer systems [1-21] have a number of known magnetic states. There is a low temperature Gd-aligned state where the Gd magnetization is aligned with the external field (and the Co is opposite) and there is a high temperature Co-aligned state where the Co magnetization is aligned with the external field (and Gd is opposite). Finally there is also a "twisted" or canted state where the moments in each atomic layer have a different angle with respect to the applied field. Therefore, Co/Gd and Fe/Gd structures are artificial ferrimagnets where the effective contribution of the two magnetic components is controllable by changing the layering pattern. At a particular temperature, which is called compensation temperature ($T_{comp}$), the magnetic moment of the Co and Gd layers are equal in magnitude and total moment goes to zero due to the antiparallel alignment.

Although there have been many investigations into the static configurations found in Fe/Gd and Co/Gd multilayers, there have been relatively few studies on the dynamical modes [22-31] found in these artificial ferrimagnets. In particular, the ferromagnetic resonance (FMR) of alternating ferromagnetic films that are antiferromagnetically coupled at the interfaces may be interesting since the magnetizations of the two materials respond very differently to temperature changes. For example, in the Co/Gd multilayer the average Gd magnetic moment changes from about 7 $\mu_B$ to zero when the temperature is varied from 0 to around 300 K, while the moment of the Co changes only slightly over the same temperature range.

Most of the earlier work on the dynamic modes [22-30] was done on FeGd and CoGd alloys and concentrated on determining the g-value, the exchange stiffness constant, the magnetic damping parameter and the anisotropy constants. An interesting property found for these amorphous alloys was the observation of multi-resonance peaks. A reasonable explanation [23] was that the non-uniformity of these amorphous alloys creates the multi peak FMR spectra since the structurally different portions of the samples may absorb the microwave energy at different fields due to the different anisotropy constants. High temperature annealing can change the multi-peak spectra of these alloys. An alternative explanation [26] was made under the assumption that the surface of these alloys can act differently than the bulk of the thin (<5000 A) film. Therefore surface and the bulk of the film can resonate at two different fields.

Our attention here focuses on the strong antiferromagnetic coupling at the interface of the Co/Gd multilayers and its effects on the FMR absorption. Furthermore we investigate the temperature dependence of the FMR signal because there are significant changes in magnetic structure with varying temperature. Indeed, our FMR results are highly sensitive to temperature changes. In fact we will see that the FMR signal for some multilayer structures simply vanishes 50 K below the room temperature, while the FMR signal for other structures can be observed down to very low temperatures.

In this paper we measure the FMR of Co/Gd multilayers, Gd/Co/Gd and Tb/Co/Tb trilayers. Our main results demonstrate substantial shifts in the resonance field due to the antiferromagnetic exchange interaction between the Co and Gd. We show that the shift in the resonant field increases as the temperature is reduced and the linewidth of the absorption peak increases as the temperature is decreased. We have also performed theoretical calculations which explain this behavior and which are in reasonable quantitative agreement with the experimental data.

Part of the motivation for this study is to produce device materials which operate at higher resonance frequencies. Such materials are of interest, for example, in high frequency signal processing [32-34] where a large operational frequency is desired at a low external magnetic field. The shift in the resonance field discussed above corresponds approximately to a 10-15 GHz boost in resonance frequency at low fields, and represents a substantial increase of the FMR frequency at low fields.

## Experimental Details

A variety of Gd/Co multilayers, Gd/Co/Gd and Tb/Co/Tb trilayers were prepared in a dc magnetron sputtering system at room temperature. The unbaked base pressure of the UHV deposition chamber was $10^{-9}$ Torr. Ultra high purity Argon gas pressure was 3 mTorr during the deposition. The samples were deposited on Corning glass substrates and 50 Å Ag layers were used as buffer and protective cap layers in all samples. The deposition thicknesses were monitored *in situ* by a quartz thickness gage which was calibrated by a stylus profilometer. Magnetization measurements were taken using a SQUID magnetometer and a Magneto-Optical Kerr Effect (MOKE) system starting from 300 K under a constant in-plane external magnetic field. The dynamic response of the samples was measured by a 24 GHz Ferromagnetic Resonance (FMR) spectrometer between 30 and 300 K. Measurements were made using a cylindrical resonant cavity which operates in the $TE_{012}$ mode. External magnetic field is parallel to the sample surface. The peak to peak linewidth was measured from the differential FMR signal.

To begin the magnetic characterization of the samples, we measured the magnetization as a function of temperature. A typical curve of the ferrimagnetic (Co 40 Å /Gd 40 Å)$_8$ multilayer is shown in Fig. 1. The results display significant thermal hysteresis, and the compensation temperature, $T_{comp}$, is around 175 K as seen from the midpoint of the "bow-tie" in Fig. 1. Details of the tunable thermal hysteresis can be found elsewhere [35-38].

We confirmed the thermal hysteresis results of Fig. 1 by measuring the MOKE signal as a function of temperature as shown in Fig. 2. Unlike the complicated thermal hysteresis results in Fig. 1, this measurement gives a nearly square hysteresis loop which might have applications in magnetic recording. The reason for the differences in Fig. 1 and Fig 2 is that the MOKE rotation angle is more sensitive to the 3d Co spins than the localized 4f Gd spins. This occurs because the MOKE signal originates from spin-orbit coupling. Since there is LS coupling in Co while in Gd there no spin-orbit contribution (L=0), the standard MOKE experiment essentially provides element specific results (Co in this case).

## Experimental Results

To understand the FMR spectra [39-40] of the Co/Gd multilayers, it is helpful to compare the results for a pure Co layer with those of the multilayers. In Fig. (3a) we present the FMR spectrum for a single 400 Å Co film at room temperature. There is a narrow peak centered at 3.5 kOe indicating an effective magnetization of 1.25 kOe for the pure Co film. This magnetization is found using the formula

$$f = \gamma\sqrt{H_{res}(H_{res} + 4\pi M_S)}\ . \qquad (1)$$

Here the gyromagnetic ratio is $\gamma$ = 2.92 GHz/kOe, $H_{res}$ is the field at which the FMR shows maximum absorption and f = 23.93 GHz is the operation frequency.

The FMR spectra for a (Co 40 Å /Gd 40 Å)$_8$ multilayer are shown in Fig. (3b-d) for different temperatures. Compared to the pure Co film, the position of the uniform FMR peak is lower by about 1 kOe at room temperature. The reason for this shift is the exchange coupling between the Co and Gd as we will discuss later on. As the temperature is lowered from the room temperature (Fig 3c and 3d), there are some small changes in the resonance field and the linewidth increases substantially. Around 250 K the FMR signal essentially vanishes.

It is not immediately clear what changes should occur as the temperature is decreased. At lower temperatures, the effective Gd moment increases in its thermal averaged magnitude and this would reduce the magnetization of the entire structure as seen in Fig. 1. Using Eq (1), the reduced magnetization would shift the resonance field to higher values. However, this can not be the only effect. As we have seen, the exchange field produced by the Gd acts on the Co and provides an effective field which dramatically lowers the resonance field of the Co/Gd multilayer compared to the pure Co film. At lower temperatures, one would expect that this exchange field would be larger and that the resonance position for the multilayer would be further lowered. For some Co/Gd multilayer samples we observed multi-peak spectra where the multiple peaks occurred above the uniform resonance field however this was not the case for every multilayer combination. These multi-peak spectra may have different sources as indicated in the introduction section. Indeed for example the curve in Fig. 3c can be better fitted by two Lorentzians numerically instead of one.

The in-plane angular dependence of the FMR field for the $(Co\ 40\ Å\ /Gd\ 40\ Å)_8$ multilayer at 24 GHz is shown in Fig. 4. The $(Co\ 40\ Å\ /Gd\ 40\ Å)_8$ multilayer can be considered as polycrystalline since it does not show any easy or hard axis anisotropies. This is consistent with the previous X-ray measurement where the $(Co\ 40\ Å\ /Gd\ 40\ Å)_8$ multilayer does not show any preferred orientations [36]. In most of our samples we did not observe any significant in-plane anisotropy.

Part of the motivation to proceed further with covering the Co with atomically thin Gd is to better focus on the coupling effects on the resonance properties. This is due to the fact that increasing Gd moment with decreasing temperature has the effect of eliminating the FMR signal as shown in Fig. 3b-d in a small (~50 K) temperature interval and this 50 K temperature interval below room temperature is not sufficiently large to process the data. Our theoretical calculations lead us to use different thickness combinations of the Co/Gd layers. Our goal was to reach large, temperature independent, shifts of the FMR field. As mentioned earlier, the motivation for the large FMR field shifts is to use this structure in microwave device applications in which the external magnetic field requirement would be substantially lower [38-40]. We therefore created trilayers of Co and Gd where the Co film (30-40 Å) is sandwiched between a few monolayers (ML) of Gd (5-7 Å) and Tb (1-7 Å). To increase the signal we repeated the trilayer structure 10 times putting a 50 Å thick nonmagnetic Ag spacer between the trilayers. The experimental results are shown in Figs. 5-7.

As shown in Fig. 5, the ferromagnetic resonance field for a 400 Å thick single film of Co is around 3.5 kOe at room temperature where resonance frequency f = 23.93 GHz. When we sandwiched the 40 Å Co film with 5 Å (~1.5 monolayers) of Gd on both sides of the Co the FMR peak shifts about 0.6 kOe lower to 2.92 kOe. The FMR peak field for $(Gd\ 7\ Å/Co\ 40\ Å/Gd\ 7\ Å/Ag\ 50\ Å)_{10}$ is slightly lower still around 2.86 kOe. This is in agreement with theoretical calculations which show that adding additional Gd layers enhances the downward shift. The $(Gd\ 5\ Å/Co\ 30\ Å/Gd\ 5\ Å/Ag\ 50\ Å)_{10}$ multilayers have a FMR peak position near 2.9 kOe. As will be discussed later, the reason for these field shifts is the strong antiferromagnetic coupling of the Co and Gd layers.

Antiferromagnetic exchange coupling acts like an extra effective field that goes into the Eq. (1) and, at a fixed frequency; this reduces the amount of the applied field which is required for the resonance.

When we repeated the same experiment with atomically thin Tb films on the outside of the 40 Å thick Co film, we could only observe an FMR signal when the Tb layer thickness was 1 Å (~0.3 monolayer). Thicker Tb films above 1 Å damped the FMR signal so that no response was visible. The resonance field shift for the (Tb 1 Å/Co 40 Å/Tb 1 Å/Ag 50 Å)$_{10}$ multilayer is on the order of 0.25 kOe from the Co peak.

The temperature dependence of the FMR field for the trilayer structures are shown in Fig. 6. The FMR field for Co is nearly constant over the entire temperature range. In contrast, the multilayers show a slight decrease in the FMR fields as the temperature is reduced. This may be due to an increased effective exchange field produced by the Gd and Tb as the temperature is reduced. Again this feature is in reasonable agreement with the theoretical results.

The FMR linewidth as a function of temperature is shown in Fig. 7 for all the samples. The linewidths generally increase monotonically as the temperatures decreases. However there is an abrupt discontinuity for the sample in Fig. 7 (top panel) below 150 K where the Gd thickness is 5 Å. This may be due to the non-uniformity of that particular sample. Although the resonance field values shown in Fig. 6 for the trilayer with 30 Å and 40 Å of Co when the Gd is 5 Å thick are similar at room temperature the gap widens at low temperatures and smaller fields for the resonance are observed when the Co is 30 Å thick. This is meaningful owing to the forthcoming theoretical discussion since the proportional amount of the Gd is slightly larger when the Co layer is thinner. However a reverse effect is shown on the linewidth pattern shown in Fig. 7 as the temperature is decreased. This time trilayer with the thicker Co deflected more as the temperature is decreased. Furthermore temperature dependent linewidth for the (Gd 5 Å/Co 30 Å/Gd 5 Å/Ag 50 Å)$_{10}$ multilayer has smaller values than that of the bare 400 Å Co film which we use as the reference film. Therefore correct comparison for the linewidth for the trilayer containing 30 Å thin Co film can be made most probably with a thinner reference Co film. Nevertheless the general features of trilayer containing 30 Å of Co with 5 Å of Gd on each side are in good agreement with other films and also with the theoretical discussion such as the downward shift of the resonance field compared to bare Co film and the monotonic decrease (increase) of the resonance field (linewidth) on the large temperature range. The FMR linewidth depends on a number of factors including the damping mechanisms and the uniformity of the structure. It is well known that Tb can significantly increase the damping [41], and indeed we find that even 1 Å of Tb on both surfaces of the Co film causes the linewidth to increase by a factor of two at room temperature and a factor of three at low temperatures. However, Gd doping in transition metals [42] does not produce such a dramatic increase. Nonetheless we find that as the temperature is lowered there is a distinct increase in the linewidth even for the Gd/Co/Gd samples. This is in agreement with the theoretical calculations and will be discussed in the next section.

**Theoretical Calculations**

There are a number of factors which need to be considered in order to understand how the resonance field measured in the Co film is influenced by the thin Gd layers. First, if the Gd spins were rigidly coupled to the Co spins by exchange coupling, one might expect either no shift in the resonance field or an upward shift in the resonance field. The reason for this is that, in this case, the Gd moments would be strictly antiparallel to the Co moments and the exchange field of the Gd would not contribute to the torque on the Co. Then the net effect of the Gd would be to reduce the net magnetization of the structure. Then using the standard formula for resonance in a thin film, Eq (1), the required field for resonance would have to be increased. If the temperature would be reduced, the net magnetization would be further reduced and the resonance field would have to be further increased. This is not what is seen experimentally.

Experimentally, one finds a <u>downward</u> shift in the resonance field and the shift is increased as the temperature is reduced, therefore the Co and Gd moments can not be strictly antiparallel. There can be several causes for this. First, the magnetization of the Gd film is very different from that in a Co film. As is well known, the precession in a thin ferromagnetic film is elliptical due to the demagnetizing fields. For example, in an isolated film the ratio of the x and y amplitudes of the magnetization when an external field H is parallel to the surfaces of the film is given by

$$\frac{M_{xo}}{M_{yo}} = \sqrt{1+\frac{4\pi M}{H}} \; . \qquad (2)$$

Since M is not the same in the Gd and Co, one would not get the same ellipticity in two materials and the moments in the two materials will not always be antiparallel. This eventually produces an effective field from the Gd onto Co, thus reducing the value of the external field H necessary to create the ferromagnetic resonance condition. These considerations agree well with the experimental results.

Now we want to consider how temperature influences the measured resonance field. It is well known that the effective exchange coupling between the Co and Gd has energy proportional to $J_I\, S_{Gd}S_{Co}$. The interface exchange constant $J_I$ is much larger than the exchange constant within Gd. As a result, the Co magnetization can stabilize the Gd moments right at the interface above the usual Curie temperature of the Gd. The calculations show that when there is only one atomic layer of Gd the ratio of the thermal averaged Gd moment at 300 K to that at 0 K is about 0.53. In a bulk Gd sample, of course, this ratio would be zero since the Curie temperature is 293 K. This explains how there can be a substantial shift in $H_{res}$ even at room temperature. Furthermore, as the temperature is reduced, the thermal averaged Gd moment increases, thereby increasing the effective field and reducing the resonance field. Again, this behavior is in good agreement with the experimental results.

We can also qualitatively understand what happens as the number of Gd layers is increased. Adding additional Gd layers also helps stabilize the Gd layer right next to the Co. For example, when there are two atomic layers of Gd the ratio of the thermal averaged Gd moment at 300 K to that at 0 K is about 0.6, compared to the 0.53 seen when there is only one atomic Gd layer. Thus one expects to see a larger shift in the resonance field. This will be evident in the theoretical results, but the experimental results are less clear on this point.

Finally we would like to understand the behavior of the FMR linewidth as a function of temperature. The linewidth measures many different factors including relaxation mechanisms and sample uniformity. In the calculations we have chosen the linewidth of the two materials to be the same, so we are investigating the question of how the linewidth is influenced by the coupling of the two materials. Without the Gd, all the Co moments can precess in a uniform motion; the effective field acting on each Co is the same. When the Gd is added the outer Co spins see a different field than the interior ones and this nonuniformity leads to an enhancement of the linewidth. As the temperature is reduced the effective exchange field becomes larger, leading to a larger nonuniformity and a larger linewidth.

Because the interfacial coupling has an energy proportional to $J_I <S_{Gd}><S_{Co}>$, one expects the strength of the effective field acting on the Co at the interface is proportional to $J_I<S_{Gd}>$. As the temperature is reduced $<S_{Gd}>$ increases and so does the effective exchange field. This increase in effective field then reduces the resonance field at low temperatures, in agreement with the experimental results.

The calculations are done using the Landau-Lifshitz-Gilbert (LLG) equations. This technique can give information on both the static and dynamic properties of magnetic multilayers. We write a set of LLG equations for each atomic layer in the multilayer

$$\frac{d\vec{S}_i}{dt} = \gamma(\vec{S}_i \times \vec{H}_i) - \frac{\alpha}{S_i}\left(\vec{S}_i \times \frac{d\vec{S}_i}{dt}\right) \quad (3)$$

Here $H_i$ is the effective field acting on the magnetization, $S_i$, in layer i. $\gamma$ is the gyromagnetic ratio and $\alpha$ is a dimensionless damping parameter. The total effective field is a sum of the external field, exchange fields, the oscillating field, and dipolar fields and is given by

$$\vec{H}_i = H_o\hat{z} + J_{i,i+1}\langle\vec{S}_{i+1}\rangle + J_{i,i-1}\langle\vec{S}_{i-1}\rangle + he^{-i\omega t}\hat{x} + \vec{h}_d \quad (4)$$

Here $H_o$ is the static magnetic field in the z direction, $J_{i,i+1}$ is the exchange coupling constant between layer i and layer i+1. $<S_i>$ is the thermal averaged magnitude of the spin in layer i. The oscillating field h has an angular frequency ω and is applied in the x direction. We assume that the main contribution to the dipole field $h_d$ is the demagnetizing field of a thin layer acting on itself; thus, $h_d = -4\pi M_y \hat{y}$, where y is the direction perpendicular to the film surface. This approximation works well in the thin film or long wavelength limit and is appropriate for FMR calculations.

One picks an initial spin configuration and integrates the coupled set of equations forward in time using a differential equation solver such as the 4th order Runge-Kutta method. Thus one obtains $S_x(t)$, $S_y(t)$ and $S_z(t)$. In order to get sensible results appropriate to FMR one has to wait for the system to come into dynamic equilibrium with the driving field. The power absorbed in a period T is then given by

$$P = \tfrac{1}{T}\int_0^T h(t)\frac{dM_y(t)}{dt}dt \qquad (5)$$

and the integral is done numerically.

The temperature dependence is included by renormalizing the magnitude of the magnetization using the Brillouin function. For example the magnetic moment for the Co is given by $<m_{Co}> = g\mu_B S_{Co} B_S(g\mu_B S_{Co} H_{eff}/kT)$, and a similar equation holds for the magnetic moment in the Gd. Here $<m_{Co}>$ is the thermal averaged magnitude of the magnetic moment in Co, and $B_s(x)$ is the Brillouin function given by

$$B_s(x) = \left(\frac{(2S+1)}{2S}\right)\coth\left(\frac{(2S+1)x}{2S}\right) - \left(\frac{1}{2S}\right)\coth\left(\frac{x}{2S}\right). \qquad (6)$$

The effective field, $H_{eff}$, is the same field that acts in Eq (4), but here the exchange field that has the dominant contribution. In fact, the exchange constants within the Gd and Co are set by requiring that the bulk Curie temperatures for the two materials are given correctly. Initially each step of the time iteration also involves a renormalization of the magnetic moments for the Co and Gd, but after about 100,000 time steps the magnitudes no longer change substantially.

The results of the calculation are present in Figs. 8-9. The key parameters for the calculation are given by $M_{Gd}(T=0)$ = 2.06 kG, $M_{Co}$ = 1.46 kG, $J_{Co}$= 9,680 kG, $J_{Gd}$ = 198 kG, and $J_I$ = -1100 kG. We assume that g = 2 in both materials and $S_{Gd}$ = 3.5 and $S_{Co}$ = 0.86. The Fig. 8 shows the calculated FMR absorption curves for a 40 Å Co film and for the same film with 1 or 2 ML of Gd on each of the outside surfaces of the Co. The results are quite close to those found experimentally. The films with the Gd on the outside exhibit a reduced resonance field and are smaller and broader than the FMR curve for the Co alone. The shift between the pure Co resonance field and the system with 1 ML of Gd on the outside is 0.59 kOe, in good agreement with the experiment. This shift has been used to obtain the value for the interface coupling constant $J_I$. It is reassuring that this value is, in fact, quite close to values based on magnetization results.

The theoretical results for how the linewidth and resonance field depend on temperature are shown in Fig. 9. We find that as T is reduced the resonance field decreases slightly, about 0.5 kOe over the entire temperature range. Again, this is in reasonable agreement with the experimental results, and indicates the strengthening of the exchange field as the temperature is reduced. The linewidth also changes with temperature, increasing substantially as the temperature is reduced. Again the values are in reasonable agreement with the experimental results and with the general discussion above.

**Summary and Conclusions**

We have performed static and dynamic measurements and also self consistent theoretical calculations for Co/Gd and Co/Tb films. Resonance field for the Co/Gd multilayers are substantially reduced compared to the pure Co film. This shift is due to the antiferromagnetic coupling between the Co and Gd. As the temperature is decreased from room temperature the effective Gd magnetic moment increases and this reduces the total magnetic moment of the Co/Gd multilayer system. We have observed the broadening of the FMR linewidth for Co/Gd trilayer and multilayers as the temperature reduces from room temperature and the Gd becomes magnetic. FMR signal vanishes approximately 50 K below room temperature for Co/Gd multilayers when the Gd layer is thicker than a few atomic layers. Therefore Gd/Co/Gd trilayers where the thickness of the Gd is 1-2 monolayers are better choice to study the temperature dependence of the antiferromagnetic coupling dynamically. These Gd doped trilayers also showed a substantial downward shift of the FMR field on the order of 600 Oe at 24 GHz compared to that of pure Co film at room temperature. Likewise multilayers of Co/Gd with thicker Gd layer shows even greater downward shifts on the order of 1 kOe compared to pure Co peak at room temperature. Our theoretical calculations relate this substantial downward shift of the FMR field to the effective field from the Gd onto the Co due to the antiferromagnetic exchange coupling between them. As the temperature decreases resonance peak position at room temperature slightly decreases and the associated linewidth increases gradually at the same time. When the same experiments are repeated for the Tb/Co/Tb trilayers the FMR signal is only observable if the Tb layer thickness is on the order of 1 Å or less. Higher thicknesses over 1 Å of Tb damp the FMR signal. The results of Tb/Co/Tb trilayers are supportive and comparable with the arguments used for the case of Gd/Co/Gd trilayers. The experimental results are in good agreement with the theoretical calculations.

**Acknowledgements**

The work at UCCS was supported by DOD Grant # W911NF-04-1-0247. The work at UTA is supported by a grant (No. Y-1215) from The Welch Foundation. The work at Miami University was supported by U.S. DOE FG02-86ER45281.

**Figure Captions**

**Fig. 1.** Thermal hysteresis with increasing external magnetic fields for the (Co 40 Å/Gd 40 Å)$_8$ multilayer.

**Fig. 2.** An example of MOKE thermal hysteresis for (Co 40 Å /Gd 40 Å)$_8$ multilayer. Cooling and heating curves are measured under a constant 150 Oe inplane magnetic field is applied. Drifts on either side of the hysteresis are artifacts from the electronics of the setup.

**Fig. 3.** FMR spectrum of a) 400 Å Co film b-d) (Co 40 Å/Gd 40 Å)$_8$ multilayer as a function of temperature.

**Fig. 4.** In plane angular dependence of FMR field for the (Co 40 Å /Gd 40 Å)$_8$ multilayer at room temperature.

**Fig. 5**. Room temperature FMR spectra for Co 400 Å (full squares), (Gd 5 Å/Co 40 Å/Gd 5 Å/Ag 50 Å)$_{10}$ (full circles), (Gd 7 Å/Co 40 Å/Gd 7 Å/Ag 50 Å)$_{10}$ (empty circles), (Gd 5 Å/Co 30 Å/Gd 5 Å/Ag 50 Å)$_{10}$ (full triangles) and (Tb 1 Å/Co 40 Å/Tb 1 Å/Ag 50 Å)$_{10}$ (empty squares) films.

**Fig. 6.** FMR field as a function of temperature for various films. For the multilayer films there is a gradual decrease in the FMR field as the temperature is reduced.

**Fig. 7.** Linewidth of the FMR field as a function of temperature for various films. The linewidth of the multilayer films generally increases as the temperature is reduced.

**Fig. 8.** Theoretical results for FMR absorption for a 40 Å Co film sandwiched between 0, 1 or 2 monolayers of Gd on each side.

**Fig. 9**. Theoretical results showing how the resonance field and the linewidth changes as a function of temperature for a 40 Å Co film and for a 40 Å Co film with one or two ML of Gd on each side.

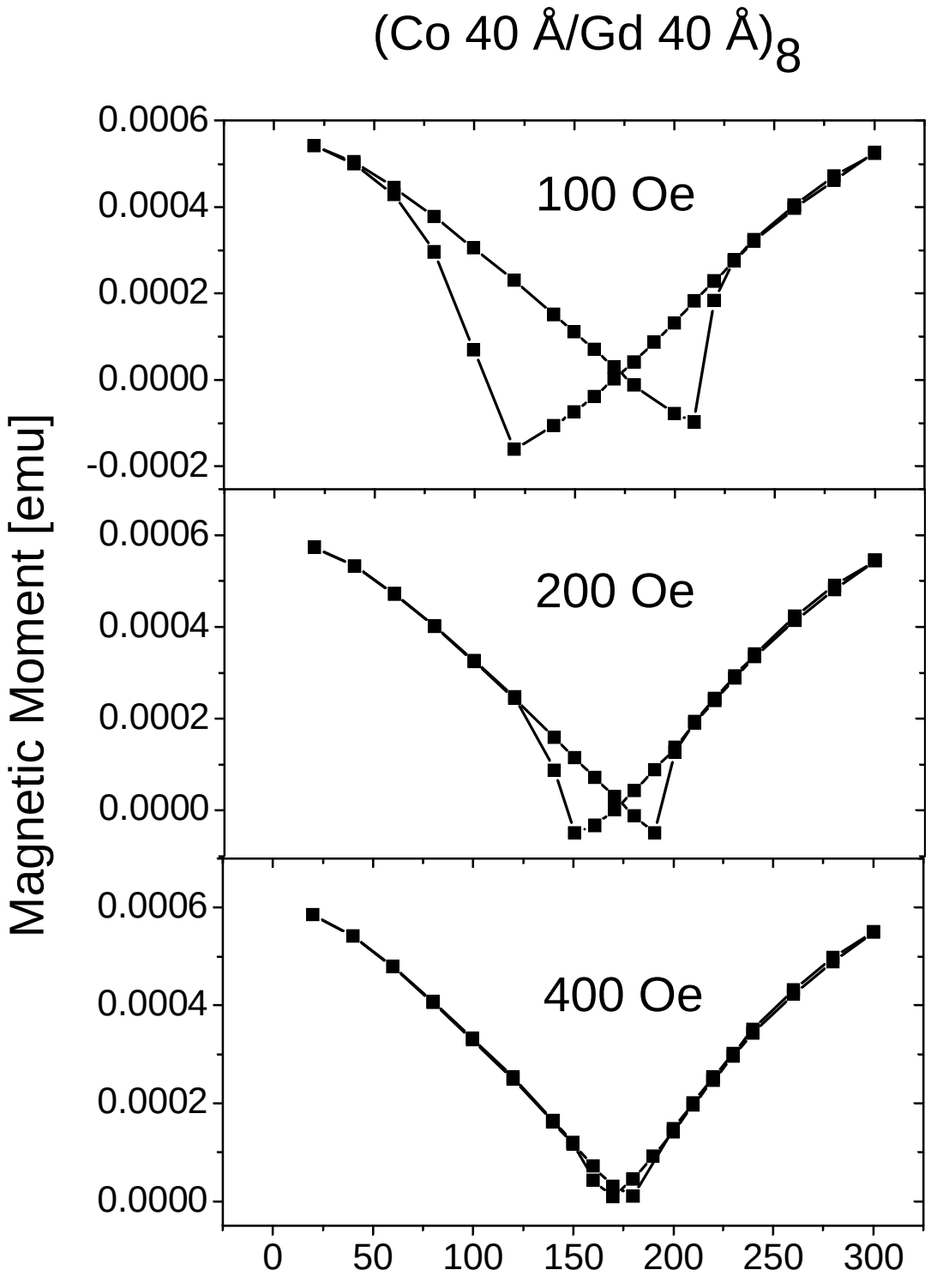

**Fig. 1.** Thermal hysteresis with increasing external magnetic fields for the (Co 40 Å/Gd 40 Å)$_8$ multilayer.

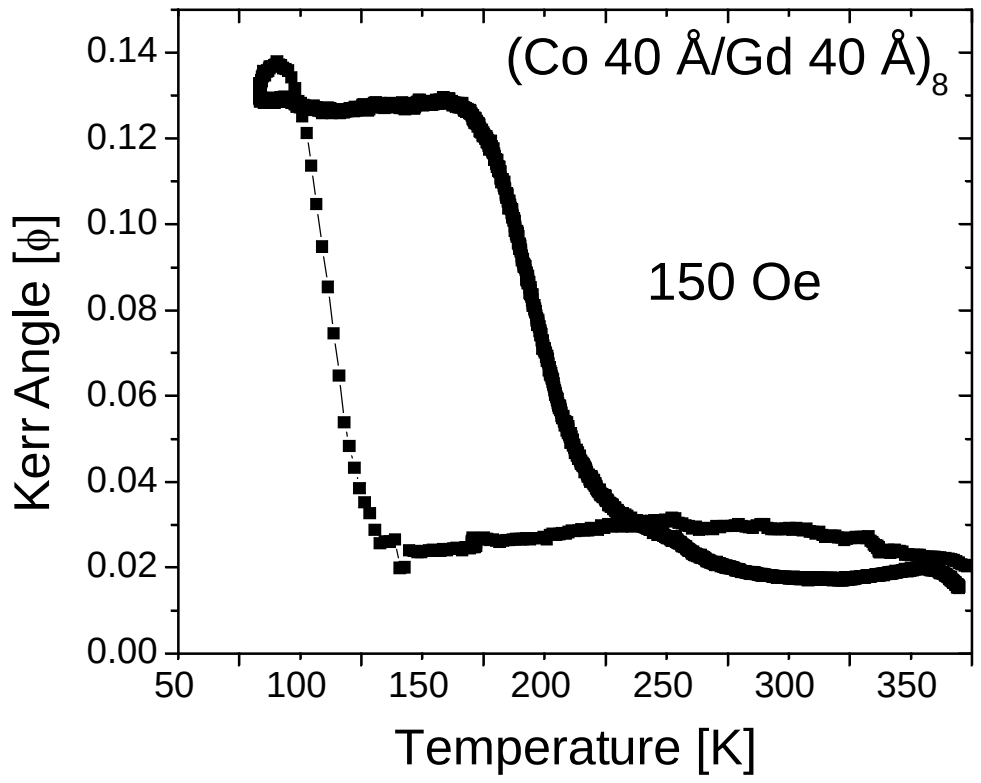

**Fig. 2.** An example of MOKE thermal hysteresis for (Co 40 Å /Gd 40 Å)$_8$ multilayer. Cooling and heating curves are measured under a constant 150 Oe inplane magnetic field is applied. Drifts on either side of the hysteresis are artifacts from the electronics of the setup.

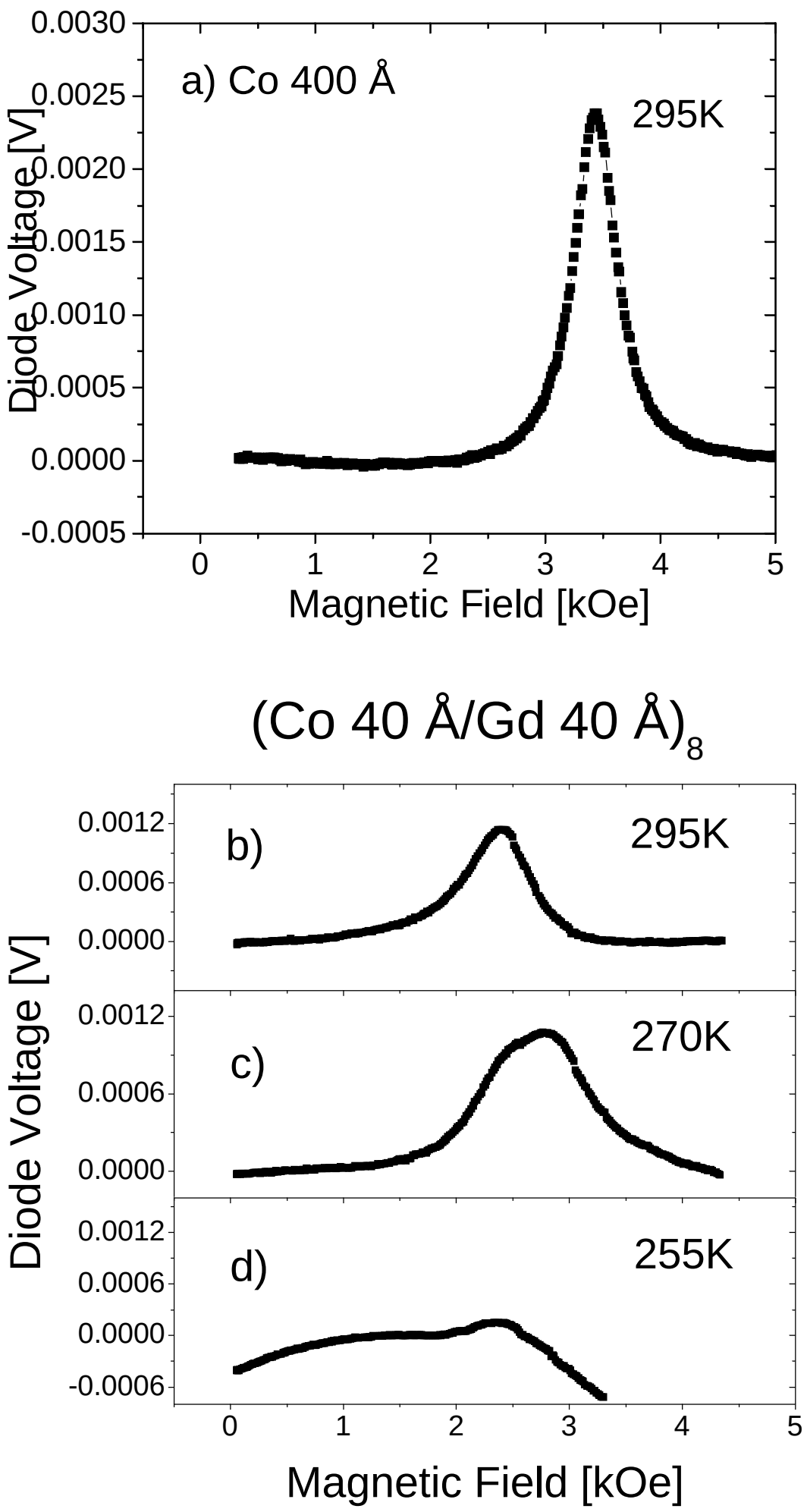

**Fig. 3.** FMR spectrum of a) 400 Å Co film b-d) (Co 40 Å/Gd 40 Å)$_8$ multilayer as a function of temperature.

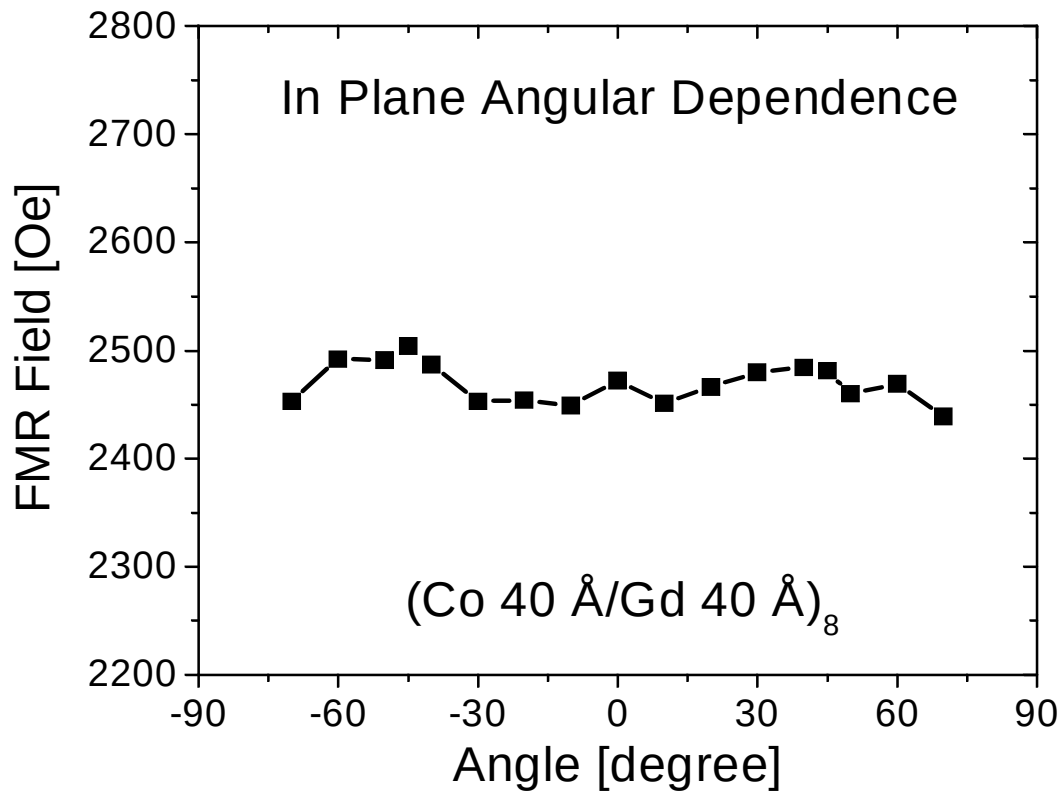

**Fig. 4.** In plane angular dependence of FMR field for the (Co 40 Å /Gd 40 Å)$_8$ multilayer at room temperature.

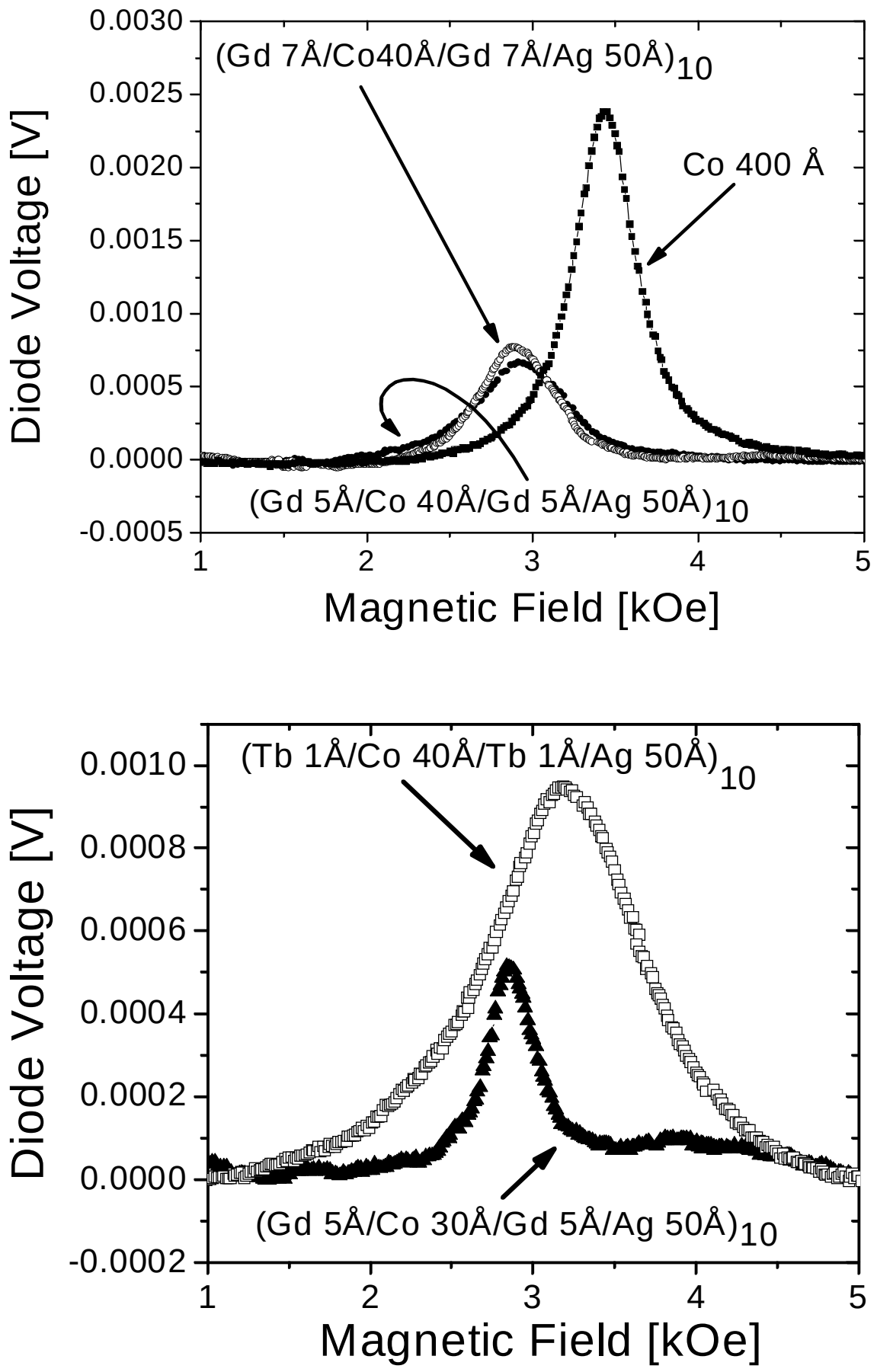

**Fig. 5**. Room temperature FMR spectra for Co 400 Å (full squares), (Gd 5 Å/Co 40 Å/Gd 5 Å/Ag 50 Å)$_{10}$ (full circles), (Gd 7 Å/Co 40 Å/Gd 7 Å/Ag 50 Å)$_{10}$ (empty circles), (Gd 5 Å/Co 30 Å/Gd 5 Å/Ag 50 Å)$_{10}$ (full triangles) and (Tb 1 Å/Co 40 Å/Tb 1 Å/Ag 50 Å)$_{10}$ (empty squares) films.

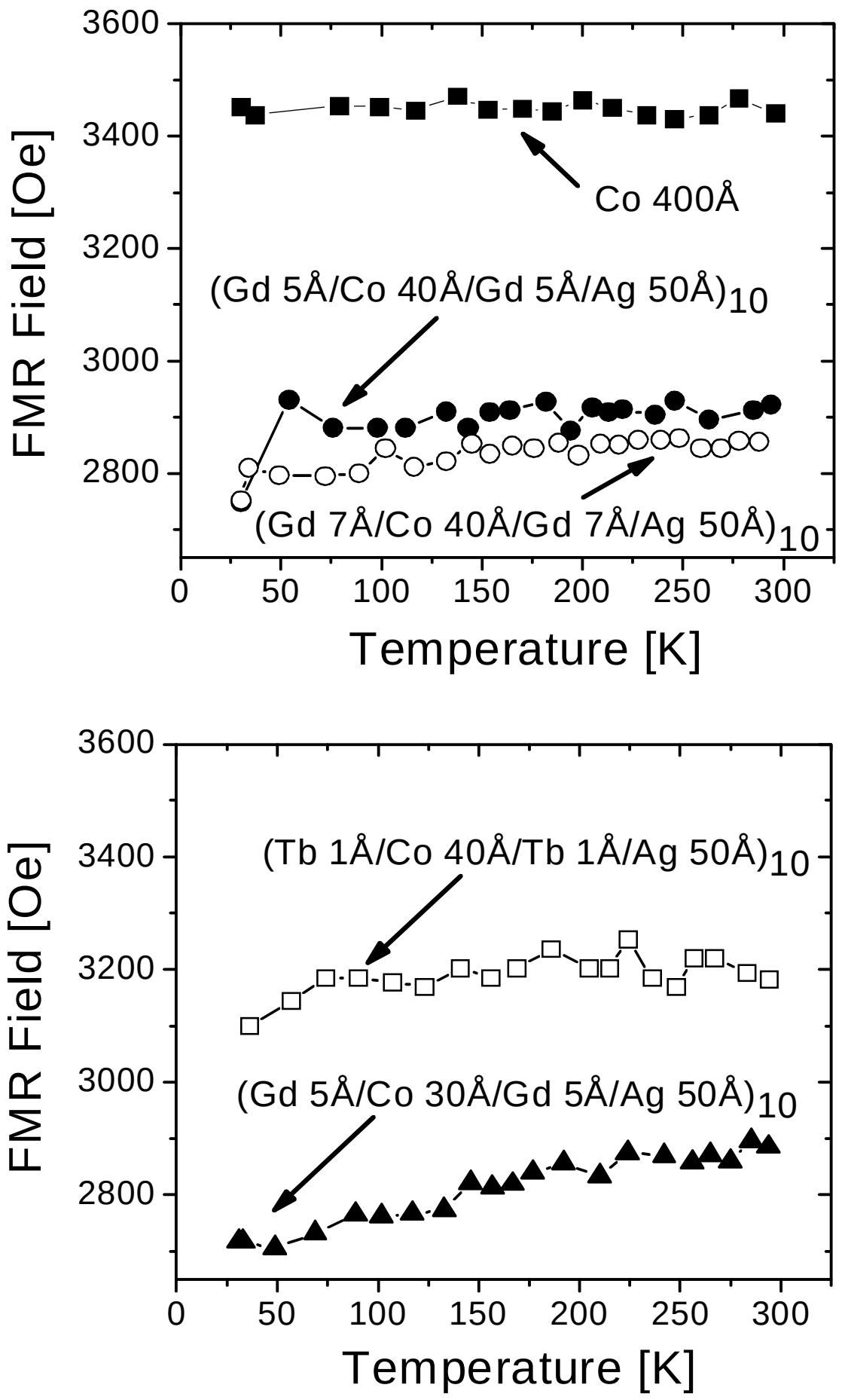

**Fig. 6.** FMR field as a function of temperature for various films. For the multilayer films there is a gradual decrease in the FMR field as the temperature is reduced.

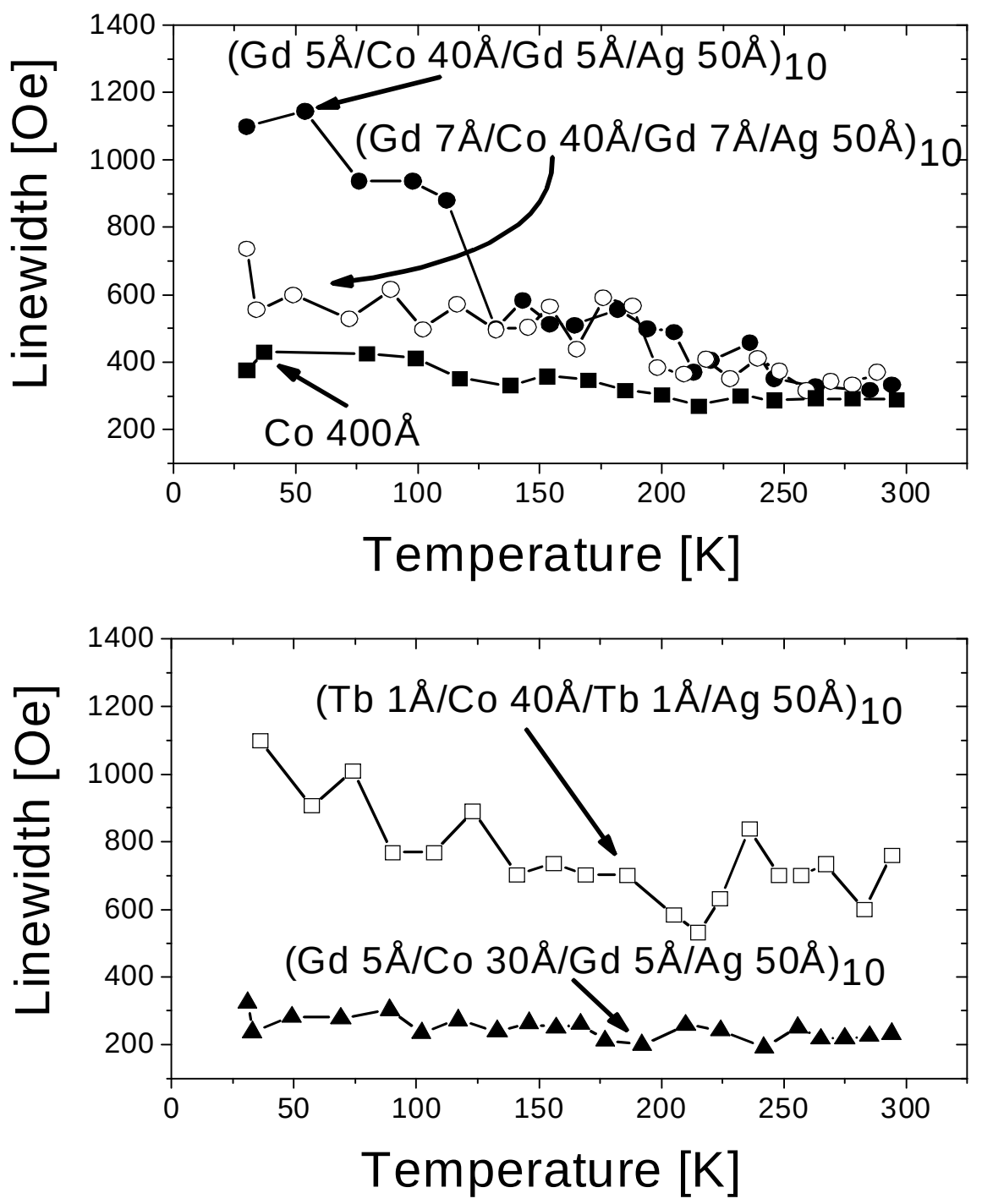

**Fig. 7.** Linewidth of the FMR field as a function of temperature for various films. The linewidth of the multilayer films generally increases as the temperature is reduced.

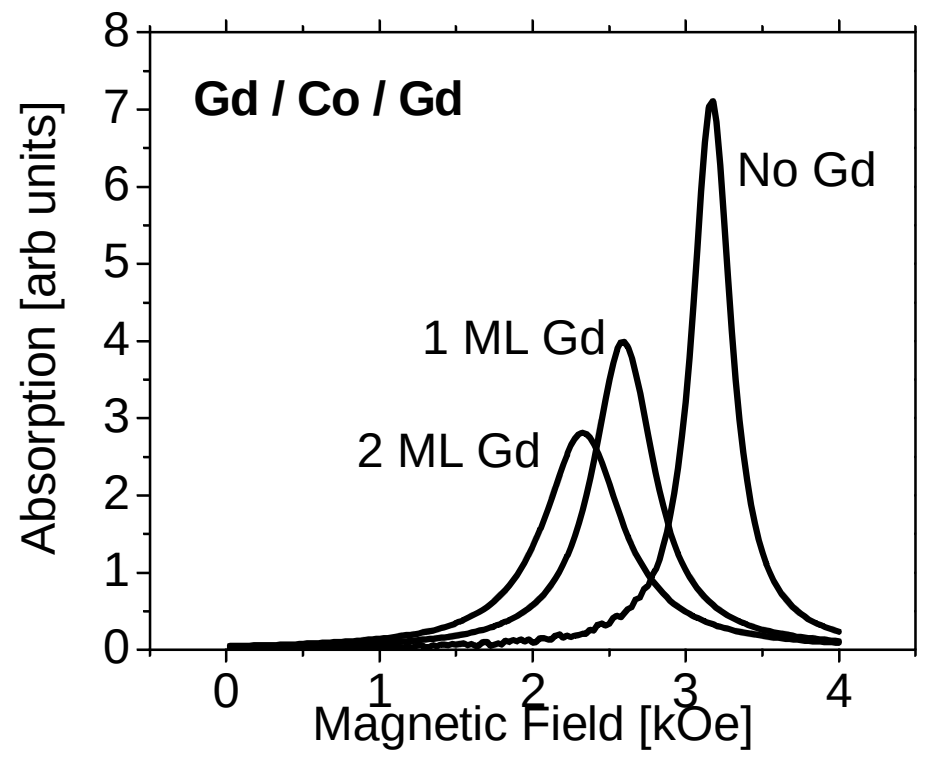

**Fig. 8.** Theoretical results for FMR absorption for a 40 Å Co film sandwiched between 0, 1 or 2 monolayers of Gd on each side.

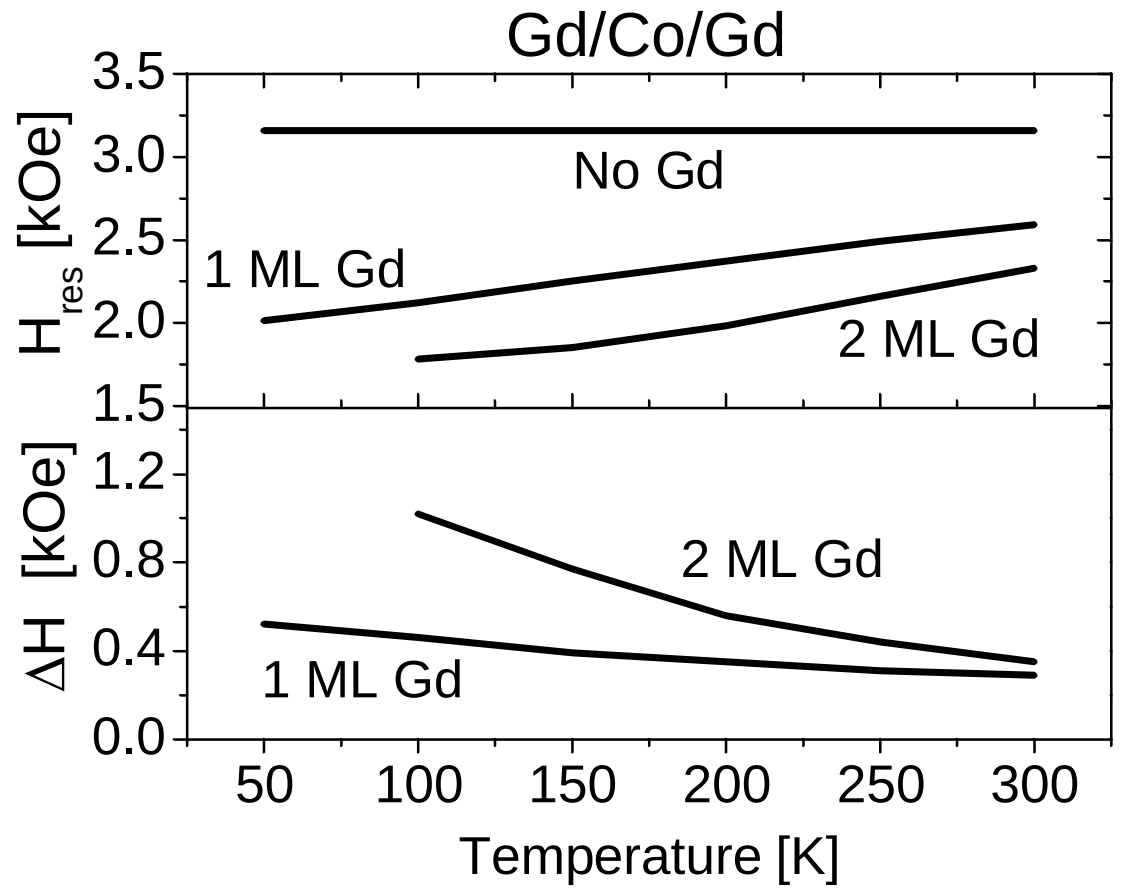

**Fig. 9**. Theoretical results showing how the resonance field and the linewidth changes as a function of temperature for a 40 Å Co film and for a 40 Å Co film with one or two ML of Gd on each side.